\documentclass[11pt, a4paper]{article}

\usepackage[utf8]{inputenc}
\usepackage{amsmath}
\usepackage{amssymb}
\usepackage{graphicx}
\usepackage{hyperref} 
\usepackage{geometry}
\usepackage{authblk}  
\usepackage{abstract}
\usepackage{booktabs} 
\usepackage{caption}
\usepackage[T1]{fontenc}
\usepackage{lmodern}

\title{Do GNN-based QEC Decoders Require Classical Knowledge? \\ Evaluating the Efficacy of Knowledge Distillation from MWPM}
\author{Ryota Ikeda}
\affil{Department of Electrical and Electronic Engineering, Yamaguchi University}
\date{August 3, 2025}

\begin{document}

\maketitle

\begin{abstract}
The performance of decoders in Quantum Error Correction (QEC) is a key factor for the realization of practical quantum computers. While Graph Neural Networks (GNNs) have emerged as a promising approach, their optimal training methodologies are not yet established. It is generally expected that transferring theoretical knowledge from classical algorithms like Minimum-Weight Perfect Matching (MWPM) to a GNN via "knowledge distillation" would be effective for performance enhancement. In this study, we test this hypothesis by rigorously comparing two models based on a Graph Attention Network (GAT) that captures temporal information as node features. The first is a purely data-driven model (baseline) trained only on ground-truth labels, and the second is a model incorporating the theoretical error probabilities from MWPM as a knowledge distillation loss. Our evaluation, using public experimental data from Google, reveals that the final test accuracy of the distillation model was nearly identical to the baseline. However, its training loss converged more slowly, and the training time increased by a factor of approximately five. This result suggests that modern GNN architectures possess a high capacity to efficiently learn complex error correlations directly from real hardware data, without guidance from approximate theoretical models.
\end{abstract}

\section{Introduction}
Quantum computers hold the potential to outperform classical computers on specific computational problems, but their realization is hindered by the fragility of qubits due to decoherence. Quantum Error Correction (QEC) is an essential technology to overcome this challenge, enabling the detection and correction of errors by redundantly encoding a single logical qubit into multiple physical qubits.

The performance of QEC is critically dependent on the classical "decoder" algorithm, which interprets the error syndrome to deduce the appropriate correction operation. The standard decoder for the surface code, Minimum-Weight Perfect Matching (MWPM) \cite{dennis2002topological}, performs well under a simplified noise model where errors are assumed to be independent and identically distributed (i.i.d.). However, noise in real quantum devices exhibits complex spatio-temporal correlations, and the discrepancy between the theoretical model and reality can degrade the decoder's performance.

To address this, decoders based on machine learning, such as Graph Neural Networks (GNNs), have emerged as a promising alternative \cite{torlai2017neural, chamberland2023graph}. GNNs have the ability to learn error patterns directly from data. It is generally anticipated that injecting physical prior knowledge into a GNN should improve its performance. Specifically, "knowledge distillation" \cite{hinton2015distilling}, which transfers the knowledge of theoretical error structures from MWPM to a GNN, is considered a concrete method to realize this hypothesis.

However, is this common expectation truly correct? For a powerful data-driven model like a GAT, does knowledge from an approximate theoretical model like MWPM genuinely aid learning, or could it potentially act as a hindrance?

In this paper, we answer this question by comparing the performance of a GNN decoder with knowledge distillation against a purely data-driven GNN decoder under completely fair conditions. Section 2 details the dataset, model architecture, and the loss function incorporating knowledge distillation. Section 3 presents the experimental results, and Section 4 discusses their implications. Finally, Section 5 provides the conclusion and future outlook.

\section{Methodology}

\subsection{Dataset and Preprocessing}
We use the public surface code experimental dataset from Google Quantum AI \cite{google2023suppressing}. From the raw measurement data (\texttt{measurements\_b8}) and sweep data (\texttt{sweep\_b8}) in the dataset, we regenerated syndrome data (\texttt{detection\_events\_new.b8}) and ground-truth labels (\texttt{obs\_flips\_actual\_new.01}) with perfectly matched shot counts using the \texttt{m2d} command from the \texttt{stim} package \cite{gidney2021stim}, ensuring data integrity. This process yielded 50,000 consistent data pairs.

\subsection{Time-Flattened Graph Construction}
To convert the time-series syndrome data into a single static graph suitable for GNNs, we introduced a "time-flattening" method. First, we identify a unique set of spatial nodes $V_s$ from the detector coordinates across all measurement rounds, ignoring the time axis. For our dataset, $|V_s|=4$. Next, we define the feature vector $\mathbf{x}_v$ for each spatial node $v \in V_s$ as a $T$-dimensional vector containing the syndrome history at that location over all $T$ measurement rounds. In our experiment, $T=2$, so each node has a 2-dimensional feature vector. The graph's edge set $E$ is constructed as a complete graph connecting all node pairs, ensuring an information propagation path exists between any two nodes.

\subsection{GNN Architecture}
The GNN model used in this study is based on the Graph Attention Network (GAT) \cite{chamberland2023graph}. The architecture consists of the following components:
\begin{enumerate}
    \item Two \texttt{GATv2Conv} layers, each followed by \texttt{LayerNorm} and a \texttt{ReLU} activation function. \texttt{GATv2Conv} provides a more expressive dynamic attention mechanism, effectively learning the interaction between node and edge features.
    \item A \texttt{global\_mean\_pool} layer to aggregate graph-wide features.
    \item A two-layer MLP classifier that outputs the final logical error probability (logit).
\end{enumerate}
This model is designed with two distinct output heads:
\begin{itemize}
    \item \textbf{Graph Classification Head:} Predicts the overall logical error for the graph from the aggregated vector after the \texttt{global\_mean\_pool}.
    \item \textbf{Edge Classification Head:} Predicts the probability (logit) that each edge is part of an error chain by concatenating the node features at both ends of the edge after the \texttt{GAT} layers.
\end{itemize}

\subsection{Loss Function and Knowledge Distillation}
\subsubsection{Data Loss ($L_{\text{data}}$)}
The dataset has a severe class imbalance (negative examples outnumber positive examples by about 11 to 1). To correct for this, we use a weighted binary cross-entropy loss. The weight for the positive class (logical error present), $w_p$, is calculated as $w_p = N_{\text{neg}} / N_{\text{pos}}$. The loss function is defined as:
\begin{equation}
    L_{\text{data}} = \text{BCEWithLogitsLoss}(y_{\text{pred}}, y_{\text{true}}, \text{pos\_weight}=w_p)
\end{equation}
where $y_{\text{pred}}$ is the output of the graph classification head and $y_{\text{true}}$ is the ground-truth label.

\subsubsection{Knowledge Distillation Loss ($L_{\text{distill}}$)}
The knowledge from the teacher model (MWPM) is extracted from the correlated error probabilities $p_{ij}$ within the \texttt{.dem} file. We define the knowledge distillation loss as the Mean Squared Error (MSE) between the teacher's probabilities $p_{ij}$ and the output of the GNN's edge classification head $\hat{p}_{ij}$ (after applying a sigmoid function to convert logits to probabilities).
\begin{equation}
    L_{\text{distill}} = \frac{1}{|\mathcal{E}|} \sum_{(i,j) \in \mathcal{E}} (\sigma(\hat{p}_{ij}) - p_{ij})^2
\end{equation}
where $\mathcal{E}$ is the set of edges in the graph and $\sigma$ is the sigmoid function. This loss term encourages the GNN to mimic the theoretical error structure understanding of MWPM.

\subsection{Experimental Setup}
In this study, we compared the following two models:
\begin{itemize}
    \item \textbf{Baseline Model:} Trained using only the $L_{\text{data}}$ loss. This model learns solely from the data without any prior physical knowledge.
    \item \textbf{Distillation Model:} Trained using a composite loss function $L_{\text{total}} = L_{\text{data}} + \lambda \cdot L_{\text{distill}}$, with the distillation weight set to $\lambda=0.5$.
\end{itemize}
To ensure a fair comparison, the architecture, optimizer (Adam, $lr=10^{-3}$), batch size (64), and \texttt{static\_edge\_weights} initialized with \texttt{random\_seed=42} were kept identical for both models.

\section{Results}
\begin{figure}[h!]
    \centering
    \includegraphics[width=\textwidth]{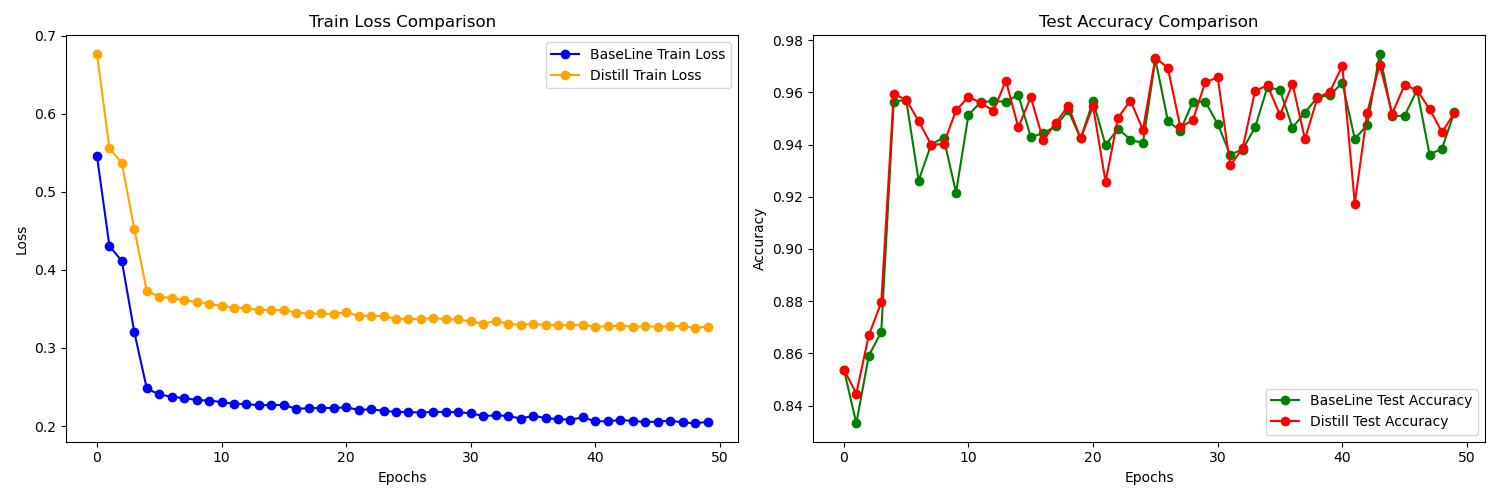} 
    \caption{Learning curves for the baseline and distillation models. The left panel shows the training loss progression, and the right panel shows the test accuracy progression. The distillation model exhibits slower convergence in training loss, but the final test accuracy is nearly identical for both models.}
    \label{fig:learning_curves}
\end{figure}

Figure \ref{fig:learning_curves} shows the learning curves for the baseline and distillation models. While the final test accuracy of both models was nearly identical, there were clear differences in the training process. The training loss of the baseline model (blue line) converged rapidly, whereas the training loss of the distillation model (orange line) consistently remained higher and showed a slower convergence.

Table \ref{tab:performance} summarizes the final performance after 50 epochs of training. There was no significant difference in the final test accuracy between the two models, while the training time for the distillation model was approximately 5.2 times longer than that of the baseline model.

\begin{table}[h!]
    \centering
    \caption{Final Performance Comparison}
    \begin{tabular}{@{}lcc@{}}
        \toprule
        Model & Final Test Accuracy (\%) & Training Time (s) \\
        \midrule
        Baseline Model ($L_{\text{data}}$ only) & 96.56 & 642.30 \\
        Distillation Model ($L_{\text{data}} + L_{\text{distill}}$) & 96.58 & 3347.93 \\
        \midrule
        Reference: Pymatching \cite{higgott2022pymatching} & \multicolumn{2}{c}{98.41\% (Error Rate 0.0159)} \\
        \bottomrule
    \end{tabular}
    \label{tab:performance}
\end{table}

\section{Discussion}
Our experimental results demonstrate that knowledge distillation from MWPM does not significantly improve the performance of the GNN decoder used in this study, but rather substantially degrades its training efficiency. This unexpected outcome suggests several important insights.

First, powerful GNNs with attention mechanisms like GAT may have a very high capacity to learn complex, higher-order error correlations directly from sufficient data—correlations that are not captured by the simplified physical models designed by humans. In our setup, it is plausible that the GNN's self-learning ability surpassed the benefits of the external "advice".

Second, the discrepancy between the theoretical model and reality may have acted as a hindrance to learning. The source of MWPM's knowledge, the \texttt{.dem} file, is ultimately an approximate model of the real noise. Attempting to instill this imperfect knowledge into the GNN may have impeded its ability to learn the true noise patterns from the real data. The slower convergence of the training loss can be interpreted as the result of the two loss terms, $L_{\text{data}}$ and $L_{\text{distill}}$, attempting to optimize the model in conflicting directions.

Furthermore, the more than fivefold increase in training time is a significant practical drawback. This additional cost arises from the need to implement an extra edge prediction head in the GNN and to compute MWPM probabilities for each batch. Without a corresponding improvement in final accuracy, this increase in computational cost is difficult to justify.

This finding serves as a cautionary note in the field of QEC decoders against the simple assumption that "adding physical information improves performance," and instead poses a deeper question: "What kind of information should be provided, and how?"

\section{Conclusion}
In this work, we have demonstrated that, under specific conditions, knowledge distillation from MWPM degrades the training efficiency of a GNN decoder without improving its final accuracy. This result highlights the strong capability of GNNs to learn error physics directly from data. Future work could involve verifying the effectiveness of knowledge distillation for different code topologies, such as QLDPC codes, or when using more faithful physical models. Exploring more sophisticated knowledge distillation frameworks, such as adaptive methods that dynamically change the distillation weight $\lambda$ during training, or techniques that reduce the distillation loss weight for samples with a large discrepancy from the theoretical model, also represent promising directions.

\section*{Acknowledgments}
This work was conducted as an independent project by the author.


\end{document}